# The Effect of Network Adoption Subsidies: Evidence from Digital Traces in Rwanda

Daniel Björkegren[*]
dan@bjorkegren.com
Brown University
Providence, RI

Burak Ceyhun Karaca
Keystone Strategy
New York City
karacaburakceyhun@gmail.com

## ABSTRACT

Governments spend billions of dollars subsidizing the adoption of different goods. However, it is difficult to gauge whether those goods are resold, or are valued by their ultimate recipients. This project studies a program to subsidize the adoption of mobile phones in one of the poorest countries in the world. Rwanda subsidized the equivalent of 8% of the stock of mobile phones for select rural areas. We analyze the program using 5.3 billion transaction records from the dominant mobile phone network. Transaction records reveal where and how much subsidized handsets were ultimately used, and indicators of resale. Some subsidized handsets drifted from the rural areas where they were allocated to urban centers, but the subsidized handsets were used as much as handsets purchased at retail prices, suggesting they were valued. Recipients are similar to those who paid for phones, but are highly connected to each other. We then simulate welfare effects using a network demand system that accounts for how each person's adoption affects the rest of the network. Spillovers are substantial: 73-76% of the operator revenue generated by the subsidy comes from nonrecipients. We compare the enacted subsidy program to counterfactual targeting based on different network heuristics.

[*]Corresponding author.

## 1 INTRODUCTION

Governments and NGOs spend substantial resources distributing subsidies and in kind transfers. A large literature considers the structure of these distributions: whether they reach intended recipients [1, 2], whether they are used [11], and how they compare to cash transfers [12]. But what we know about these programs comes primarily from surveys or administrative records, which tend to gather coarse measures using small samples. This paper studies a government handout that was unique in that the good provided automatically and anonymously recorded a rich stream of data about how, and by whom, it was ultimately used.

In particular, governments spend billions of dollars to connect poor and rural consumers to communication networks [16, 18]. A common program is to subsidize the adoption of either basic handsets, or smartphones (such as a program to provide smartphones to 5 million people in the Indian state of Chhattisgarh [17]). Subsidizing adoption can theoretically improve welfare because network goods generate spillovers [20]. Adoption benefits not only one's contacts: by influencing their adoption, it also benefits others further away in the network. But there is little evidence on the effectiveness of these programs. [16] suggest handset subsidies could be beneficial based on aggregate price elasticities. But operator groups suggest that these funds are not well spent and universal access goals would be better served by eliminating them, and lowering taxes [18].

We study an adoption program implemented by the Rwandan government that allocated 53,352 handsets to rural areas, 8% of the stock of handsets at the time in 2008. We use data from 5.3 billion transaction records from Rwanda's dominant mobile phone operator. To our knowledge, this represents the first impact evaluation of a handset adoption subsidy in a developing country.[1]

While handsets were allocated to rural areas, cell tower locations reveal that many handsets found their way to urban centers. The data has separate identifiers for handsets and accounts, making it possible to trace ownership of the handset. Many handsets are handed off from rural to urban accounts, and several pass through 'middleman' accounts that briefly use the handsets in the rural areas before passing them on. We interpret this as evidence that many handsets were resold. The records also capture how the handsets were ultimately used: the ultimate recipients of the handsets used them in a similar manner to individuals who paid market prices for their phones, suggesting the handsets were not wasted.

The welfare impacts of the program depend on how recipients' adoption affected the entire network of potential adopters. Demand for classical network goods has been difficult to quantify: a person

---

[1]In the U.S., there is a literature on the LifeLine service subsidy for rural consumers [10, 15]

may adopt after a contact adopts because the contact provides network benefits, or because they have similar traits or are exposed to similar environments. To gauge the welfare effects of the observed and counterfactual dispersals, we use the method to estimate and simulate demand for mobile phones in Rwanda developed in [6]. That paper overcomes simultaneity in consumer adoption decisions by inferring the value generated by each connection from subsequent interaction across that connection. Calls are billed by the second, so a subscriber must value a connection at least as much as the cost of calls placed across it. Variation in prices and coverage identifies the underlying utility of communication across each link. Consumers choose when to adopt, by weighing the increasing stream of utility from communicating with the network against the declining cost of handsets. In equilibrium, each individual reacts directly to any policy change, and to each other's responses, capturing effects that ripple through the network.

We use this method to determine how Rwanda's 2008 rural adoption subsidy program affected the network. We find that a substantial fraction of the subsidy's impact arises from spillover impacts on nonrecipients. The subsidy improved welfare.

We then compare the performance of the implemented program against alternatives that use full network information, and heuristics that can be implemented without network information. Switching to vouchers, where current subscribers are given a discount they pass along to their strongest unsubscribed contact, would improve welfare but would change the geographical distribution of benefits.

### Related Work

Our work builds on several literatures.

A literature in development economics considers the targeting of subsidized goods, typically to the neediest households [2].

Our paper is related to work that learns about societies from records from mobile phone networks, which have been used to predict wealth [9], and creditworthiness [8]. [22] uses this type data to predict the adoption of mobile money among phone users. Our study advances in two ways. First, we show how one can study the impact of a particular policy by combining knowledge of what happened on the ground, with rich details in these records that have not typically been examined in this literature (such as the sequence of accounts linked to each handsets). Second, we use a structural model to determine the impact of the implemented program, and counterfactual programs.

We use the Rwandan mobile network adoption model of [6], which in that papers is used to assess the impact of taxes and rural tower construction. It is related to the model of [23] which evaluates how the seeding of adoption in a videoconferencing network affects adoption within a firm.

A related literature considers how influence diffuses through a network [3–5, 13, 21], and typically seeks to maximize diffusion. However, maximizing diffusion may not be optimal if consumers do not internalize the social benefits or costs of adoption. The network demand system we use accounts for the social benefits of each node's adoption.

This study shows how passively collected data can evaluate the impact of policy, in one of the most data poor settings on earth.

## 2 BACKGROUND

Rwanda is a small, landlocked country in East Africa. It is predominantly rural; most households live off of subsistence farming. The country's experience with mobile phones is similar to that of other sub-Saharan African countries, though Rwanda is less developed: per capita consumption in 2005 was $265, while the World Bank reported a sub-Saharan African average of $545 [25] (all figures reported in real 2005 USD).

We focus on Rwanda during the period 2005-2009, during which the network was expanding. Because the Rwandan regulator restricted entry, the market during this period was concentrated: the mobile operator whose data we use held above 88% of the market, and its records reveal nearly the entirety of the country's remote communication. There are few alternatives for remote communication: the fixed line network is small (with penetration below 0.4%), and mail service is insignificant.[2] During this period, almost all phones were basic phones used primarily for calling; mobile internet and mobile money were not available.

### 2008 Adoption Subsidy Program

Like many countries, the Rwanda Utilities Regulatory Authority collects 2% of all operator revenues into a Universal Access Fund, to be used to for programs that accelerate Information and Communication Technologies (ICTs). In one of these programs, the Rwandan government in 2008 purchased 53,352 handsets (amounting to roughly 8% of the country's stock of handsets at the time) and distributed them to individuals through local governments at a reduced price. Fifteen of 30 districts participated in the program. Handsets were generally allocated to rural districts with low baseline mobile phone adoption (in participating districts, 4% of households had mobile phones, versus 12% in nonparticipating districts). Districts that were allocated handsets received enough for between 1% and 15% of households. The handsets were all the same model, the Motorola C113, which was chosen because it was low cost and had a long battery life.

Each district handled its own distribution; generally, individuals came to the district office to add their name to a list, and then received the handset a few months later. Beneficiaries were to pay a fraction of the full price of the handset ($28) through monthly repayments of $1.81, but few of these payments were made.

## 3 DATA

This project uses several data sources:[3]

**Call detail records (CDR):** We use anonymous call records from the dominant Rwandan operator, capturing nearly every call made over 4.5 years by the operator's mobile phone subscribers, growing from approximately 300,000 in January 2005 and to 1.5 million in May 2009. The data contains a list of transactions that can be represented as tuples: $(t, h, i, j, l_i, l_j, d)$, where $t$ is the timestamp, $h$ is the handset identifier, $i$ is the account placing the call (an anonymized identifier corresponding to a phone number), $j$ is the account receiving the call, $l_x$ is the location of the tower used to

---

[2]The average mail volume per person was 0.2 pieces per year in Rwanda, relative to 2.4 pieces in Kenya and 538.8 pieces in the US (Sources: National Institute of Statistics Report 2008, Communications Commission of Kenya, U.S. Postal Service 2011, U.S. Census).

[3]See [6] for more details about data sources.

#### Table 1: Allocation of Subsidized Handsets by District

| Household Properties | | Mean by District | | p-val of Diff. |
|---|---|---|---|---|
| | | Allocated Handsets | Not Allocated | |
| Rural | | 0.94 | 0.73 | 0.04 |
| Consump per capita | | $204 | $334 | 0.03 |
| Handsets alloc | Total | 3556.8 | 0 | 0.00 |
| | Per HH | 0.05 | 0.00 | 0.00 |
| Own phone | 2005 | 0.04 | 0.12 | 0.07 |
| | 2010 | 0.40 | 0.47 | 0.17 |
| | Increase | 0.36 | 0.36 | 0.76 |
| $N$ | | 15 | 15 | |

Sources: Banque Rwandaise de Developpement, NISR EICV.

transmit $x's$ end of the call, and $d$ is the duration. We use this data to identify subsidy recipients, measure how they are connected to each other and to the rest of the network, and measure their usage of the subsidized handset.[4]

**Coverage:** We create a baseline coverage map by computing the areas within line of sight of the towers operational in each month, a method suggested by the operator's network engineer. Elevation maps are derived from satellite imagery recorded by NASA [14, 19].

**Handset prices:** We create a monthly handset price index based on 160 popular models in Rwanda, weighting each model by the quantity activated on the network.

## 4 HOW WERE HANDSETS ALLOCATED?

We first assess targeting of the 2008 Rwandan rural phone subsidy.

### 4.1 Findings from administrative records

**Handsets were originally allocated to poor, rural districts.** Districts that were allocated handsets shaded in Figure 1, and their characteristics shown in Table 1. Handsets were generally allocated to rural districts with low baseline mobile phone adoption. Allocations varied significantly: those allocated handsets received enough for between 1% and 15% of households.

However, administrative data provides only a coarse understanding of the implementation of the policy. This particular policy can be investigated in much greater detail using digital traces from the objects that were given out.

### 4.2 Findings from digital traces

We infer which handsets are subsidized by the model, defining $subs(h) \in \{0, 1\}$. The particular model (Motorola C113) was otherwise rare in the country at the time, so we are able to identify beneficiaries based on receiving this model of handset during the dates of distribution. Figure 2 shows activations of this model over time, showing a spike at the time of the subsidy.[5] We consider an account as subsidized if its mode handset was the subsidized model, and it was activated during the first four months of 2008. (We omit handsets activated later because in later months it is difficult to tell if they are part of the program. These are held fixed in simulations, which will tend to lower our estimates of impact.) This gives us 41,225 accounts, 77% of the proposed allocation.

**Many handsets were activated in cities, not the rural areas they were allocated.** Figure 1 plots where handsets were allocated based on government records, as well as the number of subsidized handsets activated at each tower. While many handsets were activated in rural areas that received allocations, many also were activated in cities.

**A number of handsets passed through middlemen on their way to urban areas.** The data include signals of resale. One may see a particular handset $h$ activated by a phone number $i$ but subsequently passed along to phone number $i'$. Or, $i'$ may be a middleman who briefly uses the handset for testing, and then transfers it to an ultimate user $i''$. We define a middleman as an account $i'$ that uses two or more subsidized handsets $h$ for 20 or fewer transactions in between two other accounts. We find 624 subsidized handsets were transferred through 291 middlemen.

**Subsidized handsets that were activated were used, not wasted.** A common concern with subsidies is that goods may be allocated to consumers who do not value them. Because every transaction on handsets can be observed, we can assess this directly. Subsidy recipients use handsets in a similar manner as those who purchased phones around the same time. Table 2 shows that the ultimate recipients of subsidized handsets (column 3) use their phones less than individuals who subscribed earlier (column 1), but on par with individuals who paid retail for phones around the same time (column 2), in terms of calls, durations, and total number of contacts. (This is similar to [11]'s finding that subsidizing bednets did not affect the extent to which they are used.) Subsidized phones are ultimately used in more rural areas. Results are suggestive of a program that increased the supply of handsets in rural areas, with handsets ultimately being used by relatively typical users.

**Social networks are similar between those who received subsidies and those who purchased phones at retail.** We analyze the social network structure of recipients, as revealed by their later phone calls. We observe the communication graph $G_T$, where a directed link $ij \in G_T$ indicates that phone number $i$ has called $j$ by the last period of data, $T$ (May 2009).

Table 3 shows the properties of these edges. The first column presents the properties of edges in the entire network of subscribers. On average, 0.26 calls are placed per month, with a total duration that averages to 7.63 seconds. The average edge connects subscribers who are 30 km apart. 28% of edges only have calls during working hours. Most calls are short: 78% of edges have only had calls under 1 minute, and 51% have only had calls under 30 seconds. The second two columns restrict these measures to subsidy recipients: first, edges from recipients to any subscriber, and second, only

---

[4]Some months of data are missing; from the call records: May 2005, February 2009, and part of March 2009, and from the billing records: October 2006 and the months following August 2008.

[5]We define activation as the first transaction transmitted by the handset on the phone network.

## Figure 1: Handset Subsidy: Allocations and Activations

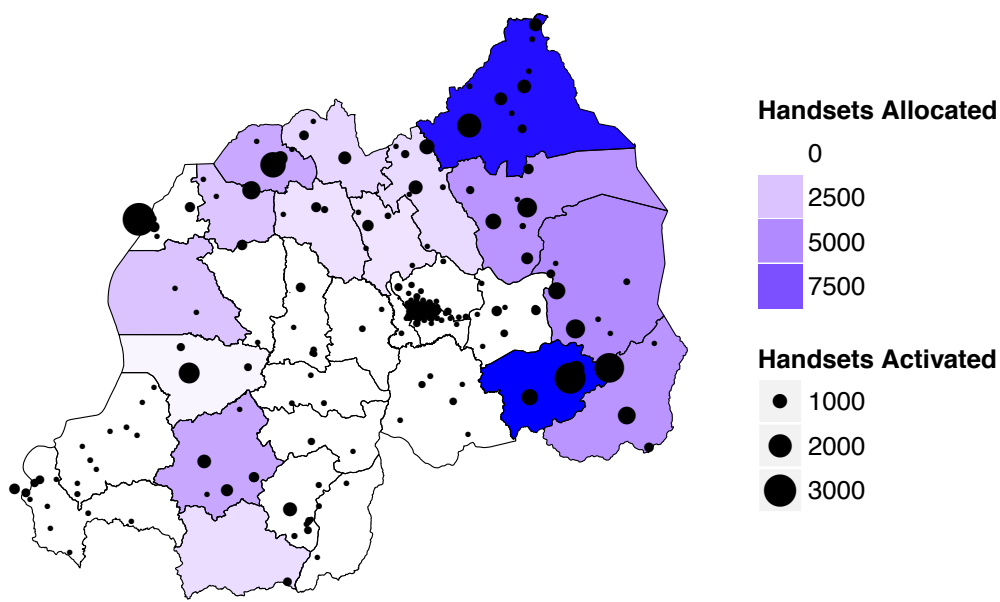

Allocations source: Banque Rwandaise de Developpement. Points represent the location of cell phone towers where subsidized handsets were first used. Motorola C113 handsets activated after January 2008 are considered distributed by the subsidy program.

## Table 2: Usage comparison

|  |  | Accounts All | Accounts Adopting 1-5.2008 | Subsidy Recipients Adopting 1-5.2008 |
|---|---|---|---|---|
| Number |  | 1,503,369 | 309,379 | 41,225 |
| Rural | Mean | 0.44 | 0.54 | 0.76 |
|  | SD | 0.50 | 0.50 | 0.43 |
| Calls | Mean | 40.0 | 37.5 | 37.7 |
| per month | Median | 24.1 | 26.1 | 28.7 |
|  | SD | 59.0 | 48.9 | 34.0 |
| Duration | Mean | 27.6 | 18.1 | 16.4 |
| minutes per month | Fraction to accounts subscribing after 1.2008 | 24% | 33% | 35% |
|  | SD | 92.2 | 47.1 | 23.0 |
| Number of Contacts (Degree) | Mean | 105.8 | 57.5 | 62.2 |
|  | SD | 159.9 | 73.4 | 42.8 |
| Clustering Coefficient | Mean | 0.068 | 0.081 | 0.082 |
|  | SD | 0.066 | 0.070 | 0.057 |

Rural is defined as an account's mode tower being located in a rural area

**Figure 2: Activations of Motorola C113**

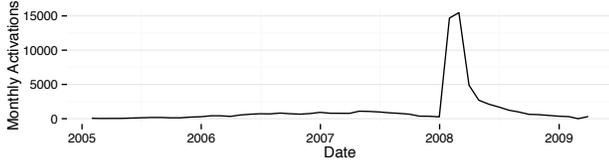

**Table 3: Edge Properties**

| Subset of Nodes: | All | Subsidy Recipients | |
|---|---|---|---|
| Edges: | All | All | Within |
| Calls per month | 0.26 | 0.32 | 0.36 |
| Duration per month | 7.63 | 7.50 | 8.02 |
| Distance (km) | 30.13 | 29.56 | 20.02 |
| Any calls during | | | |
| ...workday | 0.59 | 0.60 | 0.58 |
| ...weekend | 0.49 | 0.50 | 0.50 |
| ...late night | 0.08 | 0.07 | 0.08 |
| ...holidays | 0.28 | 0.22 | 0.21 |
| Only calls during | | | |
| ...workday | 0.28 | 0.27 | 0.24 |
| Call length | | | |
| ...all under 30 seconds | 0.51 | 0.57 | 0.55 |
| ...all under 1 minute | 0.78 | 0.84 | 0.80 |
| Nodes | 1,503,675 | 41,225 | 41,225 |
| Edges | 195.6m | 4.4m | 0.4m |

Average for the edges in the given subgraph, from January 2008 on.

edges that connect two recipients. Edges connect subsidy recipients at shorter distances (average of 20 km) but otherwise these relationships appear similar to general calling relationships.

**Subsidy recipients are connected to each other.** Among the entire population, 2% of links are with subsidy recipients. Subsidy recipients themselves are nearly 5 times more connected to each other: 9% of the links of subsidy recipients are to other subsidy recipients, as shown in the last row of Table 3.

**The people reached by the program are unlikely to generate the most network spillovers.** An optimal subsidy program to maximize network spillovers would target people who would increase the benefit of the system to others who have yet to subscribe. One measure of these benefits is the eventual duration spoken with contacts that have yet to subscribe. This measure is very similar for recipients and nonrecipients (35% for subsidy recipients, 33% for all subscribing in the same months, as shown in Table 2).[6] The fraction of a node's neighbors who are themselves connected (clustering coefficient) is very similar: 0.082 for subsidy recipients and 0.081 for all subscribing in the same months.

---

[6]Subsidy recipients represent 13% of those subscribing in these months.

## 5 IMPACTS

The ultimate impact on network adoption and welfare depends on the interaction of the recipients' adoption decision with the network of benefit flows. [6] estimates a network demand system for the Rwandan mobile phone network, using the same data. We use this structural model to simulate how equilibrium adoption would change under alternate subsidy programs, accounting for how a change to one individual's adoption affects others, which recursively affect others.

### 5.1 Model

We briefly describe the empirical model of handset adoption that we use from [6]. For more details, see that paper. The utility of owning a phone is derived from making calls, so we begin with a model of usage.

*5.1.1 Usage.* Let $S_t$ be the subset of nodes subscribing in month $t$. At each period $t$, individual $i$ can call any contact $j$ that currently subscribes, $j \in G_i^T \cap S_t$, to receive utility $u_{ijt}$. Each month, $i$ draws a communication shock $\epsilon_{ijt} \sim F_{ij}$ representing a desire to call contact $j$. Given the shock, $i$ chooses a total duration $d \geq 0$ for that month, solving:

$$u_{ijt} = \max_{d \geq 0} v_{ij}(d, \epsilon_{ijt}) - c_{ijt}d$$

where $v(d, \epsilon)$ represents the benefit of making calls.

The per-second cost of placing a call is $c_{ijt} = p_t + \beta_{coverage}\phi_{it}\phi_{jt}$, which depends on the calling price $p_t$ as well as the hassle of placing a call when coverage is imperfect ($i$'s coverage at time $t$ is $\phi_{it} \in [0,1]$, the fraction of the area surrounding his most used locations receiving coverage in month $t$).

We model the benefit of making calls as:

$$v_{ij}(d, \epsilon) = d - \frac{1}{\epsilon}\left[\frac{d^\gamma}{\gamma} + \alpha d\right]$$

where the first term represents a linear benefit and the second introduces decreasing marginal returns. $\gamma > 1$ controls how quickly marginal returns decline. $\alpha$ is a cost-dependent censoring parameter that controls the intercept of marginal utility, and thus affects the fraction of months for which no call is placed.

The expected utility $i$ receives from being able to call $j$ in time period $t$ is given by:

$$Eu_{ij}(p_t, \boldsymbol{\phi}_t) = \int_{\epsilon_{ijt}}^{\infty} \left[d(\epsilon, p_t, \boldsymbol{\phi}_t) \cdot \left(\frac{1}{\beta_{cost}}\left(1 - \frac{\alpha}{\epsilon}\right) - p_t - \beta_{coverage}\phi_{it}\phi_{jt}\right) - \frac{1}{\beta_{cost}\epsilon}\frac{d(\epsilon, p_t, \boldsymbol{\phi}_t)^\gamma}{\gamma}\right] dF_{ij}(\epsilon)$$

where $\boldsymbol{\phi}_t$ represents the vector of coverage for all individuals.

*5.1.2 Adoption.* The utility of having a phone in a given period is given by the utility of communicating with contacts that have phones: each month $i$ is on the network, he receives expected utility:

$$Eu_{it} = \sum_{j \in G_i \cap S_t} Eu_{ij}(p_t, \boldsymbol{\phi}_t) + \eta_i$$

where $u_{ijt}$ represents calls from $i$ to $j$ (which $i$ pays for and thus values by revealed preference). Individual $i$ chooses when to adopt by weighing the discounted stream of these benefits against the declining price of a handset, which is represented by the price index $p_t^{handset}$. Then, $i$ considers the utility of adopting at time $\tau$ to be:

$$U_i^\tau = \sum_{t=\tau}^\infty \delta^t E u_{it}(p_t, \phi_t) - \delta^\tau p_\tau^{handset}$$

*5.1.3 Network Adoption Equilibrium.* There is some initial set of adopters $S_0$; their decisions are held fixed. Each other individual $i$ decides on an adoption time $\tau_i \in [1, ..., \bar{T}]$ to maximize his payoff $U_i^{\tau_i}(\tau_{G_i})$, which depends on his contacts' adoption decisions ($\tau_{G_i}$).

An equilibrium corresponds with a Nash equilibrium of the game where each individual simultaneously announces their adoption date $\tau_i$ at the beginning of time (a complete information static game). Specifically, an equilibrium $\Gamma$ is defined by adoption dates $\boldsymbol{\tau} = [\tau_i]_{i \in S}$ such that the adoption date of each individual $i \in S \setminus S_0$ is optimal given their contacts' adoption dates: $\tau_i = \arg\max_\tau U_i^t(\tau_{G_i})$.

Under this model, each individuals' adoption and usage depend on the adoption decisions of his contacts, which in turn depend on the adoption decisions of her contacts, and so on. A perturbation of utility that causes one individual to change their adoption date can shift the equilibrium, inducing ripple effects through potentially the entire network.

*5.1.4 Estimation.* The parameters of the model are estimated from data in [6]. Individuals choose when to adopt a mobile phone and, if they adopt, how to use the phone. The decision to use a phone directly reveals the value of each connection, overcoming traditional issues with identifying the value of network goods solely from the decision to adopt. Specifically, call decisions identify the call shock distributions ($F_{ij}$), the shape of the utility function ($\gamma$ and $\alpha$), and how usage responds to costs (monetary $\beta_{cost}$ and the hassle of imperfect coverage $\beta_{coverage}$). The adoption decision identifies bounds on idiosyncratic preferences for having a phone, $[\underline{\eta}_i, \bar{\eta}_i]$.

The standard upper bound of the value $\bar{\eta}_i$ computed from [6] would suggest that recipients value phones so much that they would have adopted at the exact same month regardless of the subsidy. Because that bound is not very informative, and because subsidy recipients are observably similar to nonrecipients who subscribed in the same months, we consider upper bounds $\bar{\eta}_i^{comparable}$ that are set to equal this group of nonrecipients on average.[7]

*5.1.5 Simulation.* We use the iterated best response method developed in [6] to first simulate adoption and usage in the baseline environment, and then in counterfactual environments.

We make four assumptions about subsidy recipients:

- In the baseline, all eligible individuals took up the subsidy[8]

---

- Recipients did not delay adoption to wait for the subsidy
- Recipients preferred taking the subsidy at the point of adoption to purchasing any time in the following 4 years
- Each recipient made an average of 5 payments, so that the program represented a discount of $18.94 (based on conversations with the bank financing the program)

## 5.2 Actual Subsidy Program

In Table 4 we compute the baseline simulation ("with subsidy" in the table), as well as simulations where the subsidy has been removed. The first captures only the immediate effect of removing the subsidy: we allow each recipient to reoptimize their decision individually, without allowing those changes to ripple through the network ("proximal effect of removal"). The second is the equilibrium that results after all nodes have adjusted their decisions ("Total Impact of the Subsidy"); we also break out the incremental impact of the ripple effects ("additional ripple effect"). The first column shows the results for all nodes; subsequent columns show results for subsidy recipients and nonrecipients.

Because the decision to purchase a subsidized good only loosely reveals how much the recipient values it, the bounds we obtain are wide. The upper bound presents an optimistic scenario: targeted individuals would have delayed adoption by 2 months on average in the absence of the subsidy. The lower bound presents a more pessimistic scenario: targeted individuals would have delayed adoption by an average of 7 months.

We find:

**The subsidy improved welfare.** Factoring in the net present cost of the subsidy of $0.63m, it shifted the bounds on net welfare upward $2.7m (lower equilibrium) and $0.4m (upper equilibrium), a social rate of return of at least 51%.[9]

**A substantial fraction of benefits accrued to *non*recipients.** Recipients' utility increased by $0.79m (lower) or $0.63m (upper), from the combination of increased calling and direct value of the discount. Nonrecipients only received utility from increased calling, but obtained 63% of all consumer surplus in the lower equilibrium and 23% in the upper equilibrium, due to spillovers.

**It may have been profitable for the operator to finance the subsidy itself.** If in absence of the subsidy, the targeted individuals would have substantially delayed adoption, it would have been profitable for the firm to subsidize their adoption itself. If the firm had financed the subsidy, the bounds on its profits would shift upward by $0.25m in the lower equilibrium, but downward by $0.62m in the upper equilibrium.

**Nonrecipients account for most of the increase in revenue.** Nonrecipients account for 76% of the revenue from the subsidy in the low equilibrium and 73% in the comparable upper. In this setting, the operator has a near monopoly, and so would be able to internalize revenue generated by nearly all nodes on the network. Financing a subsidy would have been much less attractive for a competitor that was less able to internalize revenue from the rest of the network (related to [7]).

---

[7]For nonrecipients, the comparable upper bound $\bar{\eta}_i^{comparable}$ is set to the upper bound $\bar{\eta}_i$; for recipients it is set to a weighted average of the upper and lower bound: $\bar{\eta}_i^{comparable} = a\underline{\eta}_i + (1-a)\bar{\eta}_i$. we set the weight $a$ so that the average of $\bar{\eta}_i^{comparable}$ coincides with the average of $\bar{\eta}_i$ for nonrecipients: $a = \frac{\sum_{i \in S_{unsubs0801-0805}} \bar{\eta}_i/N_{unsubs0801-0805} - \sum_{i \in S_{subs}} \bar{\eta}_i/N_{subs}}{\sum_{i \in S_{subs}} \bar{\eta}_i/N_{subs} - \sum_{i \in S_{subs}} \underline{\eta}_i/N_{subs}}$.

[8]Given the decentralized nature of the implemented subsidy program, it is difficult to determine the entire set of individuals who were eligible. Since the subsidy was very attractive, we assume that all eligible individuals took up the subsidy and that it was valid only in the month they adopted.

[9]These results consider the portion of the subsidy allocated only to the 41,225 individuals we can clearly identify as recipients. The subsidy for the other 12,127 handsets would have represented an additional net present cost of $159,130. In the most extreme case where this value was destroyed through misallocation, this cost should be subtracted from the welfare gains.

**Table 4: Impact of Adoption Subsidy Program**

|  | All nodes | Recipients | Nonrecipients |
|---|---|---|---|
| Number | 1,503,670 | 41,225 | 1,462,445 |
|  |  |  |  |
| **Adoption Time** (mean) with subsidy | [24.10, 23.22] | [37.38, 37.38] | [23.73, 22.82] |
| Total Impact of Subsidy | -0.26, -0.06 | -6.81, -1.97 | -0.07, -0.01 |
| ... proximal effect of removal | -0.17, -0.05 | -6.32, -1.86 | -0.00, 0.00 |
| ... additional ripple effect | -0.09, -0.01 | -0.50, -0.11 | -0.07, -0.01 |
|  |  |  |  |
| **Revenue** with subsidy (million $) | [165.07, 173.49] | [0.88, 0.90] | [164.19, 172.59] |
| Total Impact of Subsidy | 0.88, 0.15 | 0.21, 0.04 | 0.67, 0.11 |
| ... proximal effect of removal | 0.43, 0.10 | 0.17, 0.03 | 0.25, 0.06 |
| ... additional ripple effect | 0.45, 0.05 | 0.04, 0.01 | 0.41, 0.04 |
|  |  |  |  |
| **Consumer Surplus** with subsidy (million $) | [243.74, 254.55] | [2.06, 2.09] | [241.68, 252.46] |
| Total Impact of Subsidy | 2.11, 0.82 | 0.79, 0.63 | 1.32, 0.19 |
| ... proximal effect of removal | 1.25, 0.74 | 0.74, 0.62 | 0.51, 0.12 |
| ... additional ripple effect | 0.86, 0.08 | 0.05, 0.01 | 0.81, 0.07 |
|  |  |  |  |
| **Government Revenue** with subsidy (million $) | [65.13, 68.11] | [0.15, 0.15] | [64.98, 67.96] |
| Total Impact of Subsidy | -0.26, -0.57 | -0.49, -0.60 | 0.22, 0.04 |
| ... proximal effect of removal | -0.43, -0.58 | -0.50, -0.60 | 0.08, 0.02 |
| ... additional ripple effect | 0.16, 0.02 | 0.02, 0.00 | 0.15, 0.02 |

Results in each cell reported for the lower bound and upper bound estimate of the equilibrium. Impacts represent the difference in these bounds. we hold fixed the adoption decision of 6 subsidized nodes that have crossed bounds for $\eta_i$. Utility and revenue reported in 2005 U.S. Dollars, discounted at a rate of 0.9 annually. Consumer surplus includes the surplus utility each individual receives from the call model through May 2009, minus the cost of holding a handset from the time of adoption until May 2009.

**Ripple effects were important.** Ripple effects account for 51% (lower equilibrium) or 33% (upper) of the effect on revenue and 41% (lower) or 10% (upper) of the effect on consumer surplus.

## 5.3 Alternative Targeting Rules

The actual subsidy program represents one of many potential targeting schemes that the government could have used. This section evaluates other schemes. We hold fixed the subsidy date (January 2008), amount, and number of nodes allocated (41,225) but vary who receives the subsidy. All schemes allocate the subsidy to nodes that did not receive the actual subsidy, that had yet to adopt by January 2008, but that had adopted by the end of the data (May 2009).[10]

We evaluate theoretical allocations that rely on network information that we observe, but which would not have been known at the time:

- *Priority* ranks nodes by a given metric, and then allocates the subsidy in order until the budget is depleted.
- *Random* selects nodes at random throughout the network.
- *Random super cluster* selects a highly concentrated cluster of subsidy recipients. It selects one node at random, all his unsubscribed eventual contacts, all their unsubscribed eventual contacts, and so on, until the budget is depleted.

---

[10]While the government could also subsidize individuals who do not appear in the data, we are unable to empirically evaluate the impact of doing so. This could bias our results downward (if nonadopters are good candidates for subsidization) or upwards (if the included adopters are better candidates for subsidization, but a government would not have been able to identify them).

We also evaluate implementable allocations:

- *Voucher* mimics a common strategy used by tech companies. A subset of current subscribers are given adoption vouchers for the subsidy amount that can be passed on. We assume that each selected subscriber gives this voucher to the contact he eventually talks the most with but who has yet to subscribe (his strongest unsubscribed link).

Table 5 reports simulation results from alternate targeting rules. We report firm revenue, consumer utility, government revenue, and their sum as net welfare, in the lower and upper bound equilibria. We also report the net effect on government handset revenue (which combines the cost of the subsidies and resulting difference in handset tax revenue). For random allocations we simulate from multiple random draws, and report the mean and standard deviation across draws. Allocations that generate high overall welfare tend to also generate high welfare for each of its components (firm revenue, consumer utility, and government revenue), so we focus on the aggregate measure.

Priority allocations are reported in rows 2-7 of Table 5. These use information based on how individuals eventually use the network, which would not have been available at the time. Allocations to individuals with more connections result in higher net welfare than to those with fewer, and outperform the implemented subsidy. A different metric is the fraction of an individual's contacts who have adopted - perhaps those with many latent contacts can induce more to join the network. This turns out not to be the case–there is little difference between allocating to those who have a high fraction of contacts that have adopted, and those who have a low fraction.

Random allocations are reported in rows 8-15 of Table 5. They outperform the baseline subsidy in the lower equilibrium, but are outperformed in the higher equilibrium; apart from the random allocation to rural individuals, which is outperformed in both equilibria.

However, vouchers substantially outperform all other allocations, shown in the last rows of Table 5. We evaluate vouchers given to recent adopters (adopting December 2007) and early adopters (adopting January 2005 or prior). Relative to the baseline subsidy, vouchers given to early adopters improve net welfare by $3.55m in the low equilibrium and $1.18m in the comparable high equilibrium, nearly doubling the welfare benefits of the subsidy.

We also consider restricting these allocations to rural or urban individuals. Urban allocations tend to result in higher net welfare, which likely results from correlations with network structure, but may undermine the government's distributional aims. Next, we consider the geographic distribution of welfare.

## 5.4 Distribution of Welfare

Targeting regimes affect net welfare, and also how that welfare is distributed around the country. Figure 3 shows the geographic distribution of consumer surplus resulting from the different subsidies, including net consumer surplus from calls (omitting the benefit of obtaining a handset per se). Although the implemented subsidy targets rural areas, much of its benefits go to urban centers. We compare the implemented subsidy to counterfactual voucher subsidies to early adopters in the following two plots. Relative to the implemented subsidy, providing vouchers to early adopters in urban centers (Figure 3b) would raise net welfare but redistribute surplus from rural districts in the east, north, and southwest towards cities and the center. In contrast, providing vouchers to early adopters in rural areas (Figure 3c) would have more mixed effects, and improve surplus in parts of the east.

## 6 CONCLUSION

This project studies a mobile phone handset subsidy in Rwanda. Using transaction records after the subsidized handsets has been activated, we find that recipients are very connected to each other. We then use a structural model to simulate the effect of the enacted subsidy program, and find that it improved welfare on net. We then compare the effect of the actual dispersal to that of counterfactual dispersals.

This type of analysis can make policies and behaviors legible to centralized authorities [24], which can improve policymaking, but entails new risks of surveillance. Managing these risks will require society to have deep, informed conversations about what can be measured, and how it should be used.

## ACKNOWLEDGMENTS

Daniel thanks Michael Kremer, Greg Lewis, and Ariel Pakes for guidance and encouragement, and Sendhil Mullainathan for helpful conversations. In Rwanda, I thank the staff of my telecom partner, RURA, RDB, CGIS-NUR, NISR, Minicom, PSF, EWSA, and BRD for helpful conversations and access to data. This work was supported by the Stanford Institute for Economic Policy Research through the Shultz Fellowship in Economic Policy.

**Figure 3: Geographic Distribution of Consumer Surplus
(a) Implemented Subsidy Relative to No Subsidy:**

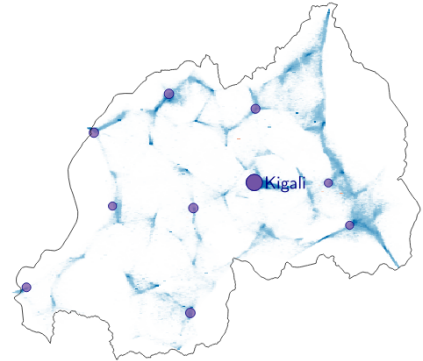

**Relative to Implemented Subsidy:**

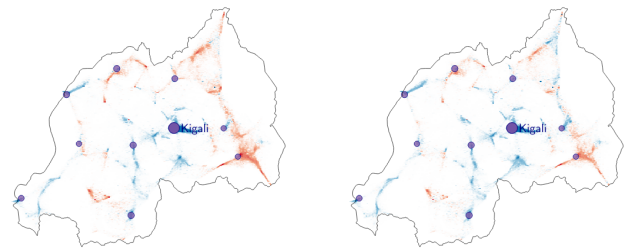

(b) Voucher to early adopters, urban

(c) Voucher to early adopters, rural

Plots show the consumer surplus derived from calls as a result of the subsidy program. Blue represents higher consumer surplus than the comparison. In the bottom two plots, red represents lower consumer surplus than the implemented subsidy.

## REFERENCES

[1] Vivi Alatas, Abhijit Banerjee, Rema Hanna, Benjamin A. Olken, Ririn Purnamasari, and Matthew Wai-Poi. 2016. Self-Targeting: Evidence from a Field Experiment in Indonesia. *Journal of Political Economy* 124, 2 (March 2016), 371–427. https://doi.org/10.1086/685299

[2] Vivi Alatas, Abhijit Banerjee, Rema Hanna, Benjamin A. Olken, and Julia Tobias. 2012. Targeting the Poor: Evidence from a Field Experiment in Indonesia. *American Economic Review* 102, 4 (June 2012), 1206–1240. https://doi.org/10.1257/aer.102.4.1206

[3] Abhijit Banerjee, Arun G. Chandrasekhar, Esther Duflo, and Matthew O. Jackson. 2013. The Diffusion of Microfinance. *Science* 341, 6144 (July 2013), 1236498. https://doi.org/10.1126/science.1236498

[4] Abhijit Banerjee, Arun G. Chandrasekhar, Esther Duflo, and Matthew O. Jackson. 2014. *Gossip: Identifying Central Individuals in a Social Network*. Working Paper 20422. National Bureau of Economic Research. https://doi.org/10.3386/w20422

[5] Lori Beaman, Ariel BenYishay, Jeremy Magruder, and Ahmed Mushfiq Mobarak. 2016. Can Network Theory-based Targeting Increase Technology Adoption? *Working Paper* (2016). https://www.atai-research.org/can-network-theory-based-targeting-increase-technology-adoption/

[6] Daniel Björkegren. 2018. The Adoption of Network Goods: Evidence from the Spread of Mobile Phones in Rwanda. *Review of Economic Studies* (2018).

[7] Daniel Björkegren. 2018. Competition in Network Industries: Evidence from the Rwandan Mobile Phone Network. *Working Paper* (2018).

[8] Daniel Björkegren and Darrell Grissen. 2019. Behavior Revealed in Mobile Phone Usage Predicts Credit Repayment. *The World Bank Economic Review* (2019). https://doi.org/10.1093/wber/lhz006

[9] Joshua Blumenstock, Gabriel Cadamuro, and Robert On. 2015. Predicting poverty and wealth from mobile phone metadata. *Science* 350, 6264 (Nov. 2015), 1073–1076. http://www.sciencemag.org/content/350/6264/1073

[10] Mark Burton, Jeffrey Macher, and John W Mayo. 2007. Understanding Participation in Social Programs: Why Don't Households Pick up the Lifeline? *The B.E.*

## Table 5: Impact of Alternate Targeting Policies

| Targeting Policy | | Welfare Net | Revenue Firm | Utility Net | Gov. Revenue All | Handset |
|---|---|---|---|---|---|---|
| Implemented Subsidy | | [3.87, 1.11] | [1.22, 0.38] | [2.74, 1.18] | [-0.08, -0.45] | [-0.45, -0.56] |
| **Theoretical Allocations** | | | | | | |
| *Deterministic* | | | | | | |
| Priority degree high | | [7.55, 1.33] | [2.39, 0.45] | [4.78, 1.19] | [0.39, -0.31] | [-0.33, -0.45] |
| Priority degree low | | [0.78, 0.20] | [0.27, 0.08] | [0.83, 0.54] | [-0.32, -0.42] | [-0.41, -0.45] |
| Priority connected to many adopters | | [7.63, 1.24] | [2.40, 0.42] | [4.85, 1.14] | [0.39, -0.32] | [-0.33, -0.45] |
| Priority connected to few adopters | | [0.82, 0.23] | [0.29, 0.09] | [0.85, 0.56] | [-0.32, -0.42] | [-0.40, -0.45] |
| Priority high fraction of connections have adopted | | [3.96, 0.83] | [1.20, 0.27] | [2.80, 0.93] | [-0.03, -0.37] | [-0.39, -0.45] |
| Priority low fraction of connections have adopted | | [3.10, 0.75] | [1.02, 0.27] | [2.14, 0.84] | [-0.06, -0.36] | [-0.37, -0.44] |
| *Random* | | | | | | |
| Random | mean | [4.08, 0.89] | [1.29, 0.30] | [2.76, 0.94] | [0.02, -0.36] | [-0.37, -0.45] |
| | sd | [0.07, 0.02] | [0.02, 0.01] | [0.04, 0.01] | [0.00, 0.00] | [0.00, 0.00] |
| Random rural | mean | [3.10, 0.66] | [1.05, 0.24] | [2.11, 0.79] | [-0.06, -0.37] | [-0.37, -0.45] |
| | sd | [0.06, 0.04] | [0.02, 0.01] | [0.04, 0.02] | [0.00, 0.00] | [0.00, 0.00] |
| Random, urban | mean | [5.57, 0.97] | [1.66, 0.31] | [3.78, 1.01] | [0.13, -0.36] | [-0.37, -0.45] |
| | sd | [0.25, 0.09] | [0.07, 0.03] | [0.16, 0.06] | [0.02, 0.01] | [0.00, 0.00] |
| Random super cluster | mean | [5.60, 0.88] | [1.77, 0.30] | [3.66, 0.94] | [0.17, -0.36] | [-0.36, -0.45] |
| | sd | [0.19, 0.07] | [0.05, 0.02] | [0.13, 0.04] | [0.01, 0.01] | [0.00, 0.00] |
| **Practical Allocations** | | | | | | |
| Voucher to early adopters | | [7.42, 2.29] | [2.28, 0.76] | [4.78, 1.73] | [0.36, -0.20] | [-0.32, -0.42] |
| Voucher to early adopters, rural | | [7.06, 2.21] | [2.19, 0.74] | [4.53, 1.67] | [0.34, -0.20] | [-0.32, -0.42] |
| Voucher to early adopters, urban | | [7.71, 2.23] | [2.34, 0.74] | [4.99, 1.70] | [0.38, -0.21] | [-0.32, -0.43] |
| Voucher to recent adopters | | [6.78, 2.13] | [2.13, 0.72] | [4.33, 1.62] | [0.32, -0.21] | [-0.31, -0.42] |
| Voucher to recent adopters, rural | | [6.16, 1.95] | [1.97, 0.67] | [3.92, 1.50] | [0.27, -0.22] | [-0.32, -0.42] |
| Voucher to recent adopters, urban | | [7.59, 2.13] | [2.31, 0.70] | [4.90, 1.65] | [0.37, -0.22] | [-0.32, -0.43] |

Results in each cell reported for the lower bound and comparable upper bound estimate of the equilibrium. Impacts represent the difference in these bounds. Utility and revenue reported in 2005 U.S. Dollars, discounted at a rate of $\delta$. Consumer surplus includes the surplus utility each individual receives from the call model through May 2009, minus the cost of holding a handset from the time of adoption until May 2009.

*Journal of Economic Analysis & Policy* 7, 1 (2007). https://doi.org/10.2202/1935-1682.1583

[11] Jessica Cohen and Pascaline Dupas. 2008. Free Distribution or Cost-Sharing? Evidence from a Malaria Prevention Experiment. *National Bureau of Economic Research Working Paper Series* No. 14406 (Oct. 2008). http://www.nber.org/papers/w14406

[12] Jesse Cunha, Giacomo De Giorgi, and Seema Jayachandran. 2017. The Price Effects of Cash Versus In-Kind Transfers. (2017).

[13] Pedro Domingos and Matt Richardson. 2001. Mining the network value of customers. In *Proceedings of the seventh ACM SIGKDD international conference on Knowledge discovery and data mining (KDD '01)*. Association for Computing Machinery, San Francisco, California, 57–66. https://doi.org/10.1145/502512.502525

[14] Tom G. Farr, Paul A. Rosen, Edward Caro, Robert Crippen, Riley Duren, Scott Hensley, Michael Kobrick, Mimi Paller, Ernesto Rodriguez, Ladislav Roth, David Seal, Scott Shaffer, Joanne Shimada, Jeffrey Umland, Marian Werner, Michael Oskin, Douglas Burbank, and Douglas Alsdorf. 2007. The Shuttle Radar Topography Mission. *Reviews of Geophysics* 45, 2 (May 2007). https://doi.org/10.1029/2005RG000183

[15] Christopher Garbacz and Herbert G. Thompson. 1997. Assessing the Impact of FCC Lifeline and Link-Up Programs on Telephone Penetration. *Journal of Regulatory Economics* 11, 1 (Jan. 1997), 67–78. https://doi.org/10.1023/A:1007902329324

[16] Christopher Garbacz and Herbert G. Thompson Jr. 2005. Universal telecommunication service: A world perspective. *Information Economics and Policy* 17, 4 (Oct. 2005), 495–512. https://doi.org/10.1016/j.infoecopol.2005.03.001

[17] Dipankar Ghose. 2018. Chhattisgarh: 50 lakh people to get smartphones under Sanchar Kranti Yojana. *The Indian Express* (July 2018). https://indianexpress.com/article/india/dual-sim-smartphones-for-50-lakh-chhattisgarh-beneficiaries-under-sanchar-kranti-yojana-5280052/

[18] GSMA. 2013. *Universal Service Fund Study.* Technical Report. GSM Association.

[19] A. Jarvis, H.I. Reuter, A. Nelson, and E. Guevara. 2008. Hole-filled seamless SRTM data V4. *CIAT* (2008). http://srtm.csi.cgiar.org

[20] Michael L. Katz and Carl Shapiro. 1994. Systems Competition and Network Effects. *The Journal of Economic Perspectives* 8, 2 (April 1994), 93–115. http://www.jstor.org/stable/2138538

[21] David Kempe, Jon Kleinberg, and Éva Tardos. 2003. Maximizing the Spread of Influence Through a Social Network. In *Proceedings of the Ninth ACM SIGKDD International Conference on Knowledge Discovery and Data Mining (KDD '03)*. ACM, New York, NY, USA, 137–146. https://doi.org/10.1145/956750.956769

[22] Muhammad R. Khan and Joshua E. Blumenstock. 2016. Predictors without Borders: Behavioral Modeling of Product Adoption in Three Developing Countries. In *Proceedings of the 22nd ACM SIGKDD International Conference on Knowledge Discovery and Data Mining (KDD '16)*. Association for Computing Machinery, San Francisco, California, USA, 145–154. https://doi.org/10.1145/2939672.2939710

[23] Stephen P. Ryan and Catherine Tucker. 2012. Heterogeneity and the dynamics of technology adoption. *Quantitative Marketing and Economics* 10, 1 (March 2012), 63–109. http://link.springer.com.ezp-prod1.hul.harvard.edu/article/10.1007/s11129-011-9109-0

[24] James C. Scott. 1998. *Seeing Like a State: How Certain Schemes to Improve the Human Condition Have Failed.* Yale University Press.

[25] WDI. 2013. World Development Indicators. *World Bank* (2013).